\pdfoutput=1
\documentclass[11pt]{article}

\usepackage[]{EMNLP2023}

\usepackage{times}
\usepackage{latexsym}
\usepackage{amsmath}
\usepackage{booktabs}
\usepackage{amsfonts}
\usepackage{makecell}
\usepackage{hyperref}
\usepackage[T1]{fontenc}
\usepackage[utf8]{inputenc}

\usepackage{microtype}

\usepackage{inconsolata}
\usepackage{xcolor}
\usepackage{wasysym}
\newcommand{\datasetname}{\textsc{PlotRetrieval }}
\newcommand{\datasetnamens}{\textsc{PlotRetrieval}}

\newcommand{\taskname}{\textit{Plot Retrieval }}
\newcommand{\tasknamens}{\textit{Plot Retrieval}}

\usepackage[]{todonotes}

\title{Plot Retrieval as an Assessment of Abstract Semantic Association}

\author{%
Shicheng Xu\thanks{\,\,Work done during the Tencent Rhino-bird Research Elite Program at WeChat.}$\,\,^{1,2}$ \quad Liang Pang$^{1}$\thanks{\ \ Corresponding authors.} \quad Jiangnan Li$^{2}$ \quad Mo Yu$^{2}$\footnotemark[2]\\ 
\bf Fandong Meng$^{2}$  \quad Huawei Shen$^{1}$  \quad Xueqi Cheng$^{1}$ \quad Jie Zhou$^{2}$ \\
$^{1}$CAS Key Laboratory of AI Security,
 Institute of Computing Technology, CAS \\
 $^{2}$Pattern Recognition Center, WeChat AI \\
  \small{\texttt{xushicheng21s@ict.ac.cn} \quad\texttt{pangliang@ict.ac.cn} \quad \texttt{{moyumyu@global.tencent.com}}}}

\begin{document}
\maketitle
\begin{abstract}

Retrieving relevant plots from the book for a query is a critical task, which can improve the reading experience and efficiency of readers. Readers usually only give an abstract and vague description as the query based on their own understanding, summaries, or speculations of the plot, which requires the retrieval model to have a strong ability to estimate the abstract semantic associations between the query and candidate plots. However, existing information retrieval (IR) datasets cannot reflect this ability well. In this paper, we propose \datasetname, a labeled dataset to train and evaluate the performance of IR models on the novel task \textit{Plot Retrieval}. Text pairs in \datasetname have less word overlap and more abstract semantic association, which can reflect the ability of the IR models to estimate the abstract semantic association, rather than just traditional lexical or semantic matching. Extensive experiments across various lexical retrieval, sparse retrieval, dense retrieval, and cross-encoder methods compared with human studies on \datasetname show current IR models still struggle in capturing abstract semantic association between texts. \datasetname can be the benchmark for further research on the semantic association modeling ability of IR models.

\end{abstract}

\section{Introduction}

% \Mo{\begin{itemize}
%     \item We propose a new task, \tasknamens, which retrieves and locates the relevant plots from book for a query.
%     \item The task is a spontaneous process in humans' daily lives. Humans naturally require to  As a result, \taskname is a common and natural scenario, but has not been well-studied in NLP. 
%     \item when reading the book or come across other life events that reminder a plot, but is not very clear. For example ...
%     \item Describe the example in the teaser
%     \item The key challenge, as shown in these examples, is the discrepancy between the query and the texts describing the target plot. Semantically associated but diverse in surface forms.
% \end{itemize}}

We propose a new task, \tasknamens, which retrieves the relevant plots from the book for a query. The task is a spontaneous process in humans' daily lives. When reading a book or coming across other life events that remind a plot, humans naturally require to find the target plot. As a result, \taskname is a common and natural scenario but has not been well-studied in NLP. 

% We propose a novel, critical and challenging task called \textit{Plot Retrieval}. \textit{Plot Retrieval} retrieves and locates the relevant plots from book for a query. It can help the readers to quickly find the part they want to read or exploit, which improves reading experience and efficiency of readers. 
\begin{figure}[t]
\centering
\includegraphics[width=\linewidth]{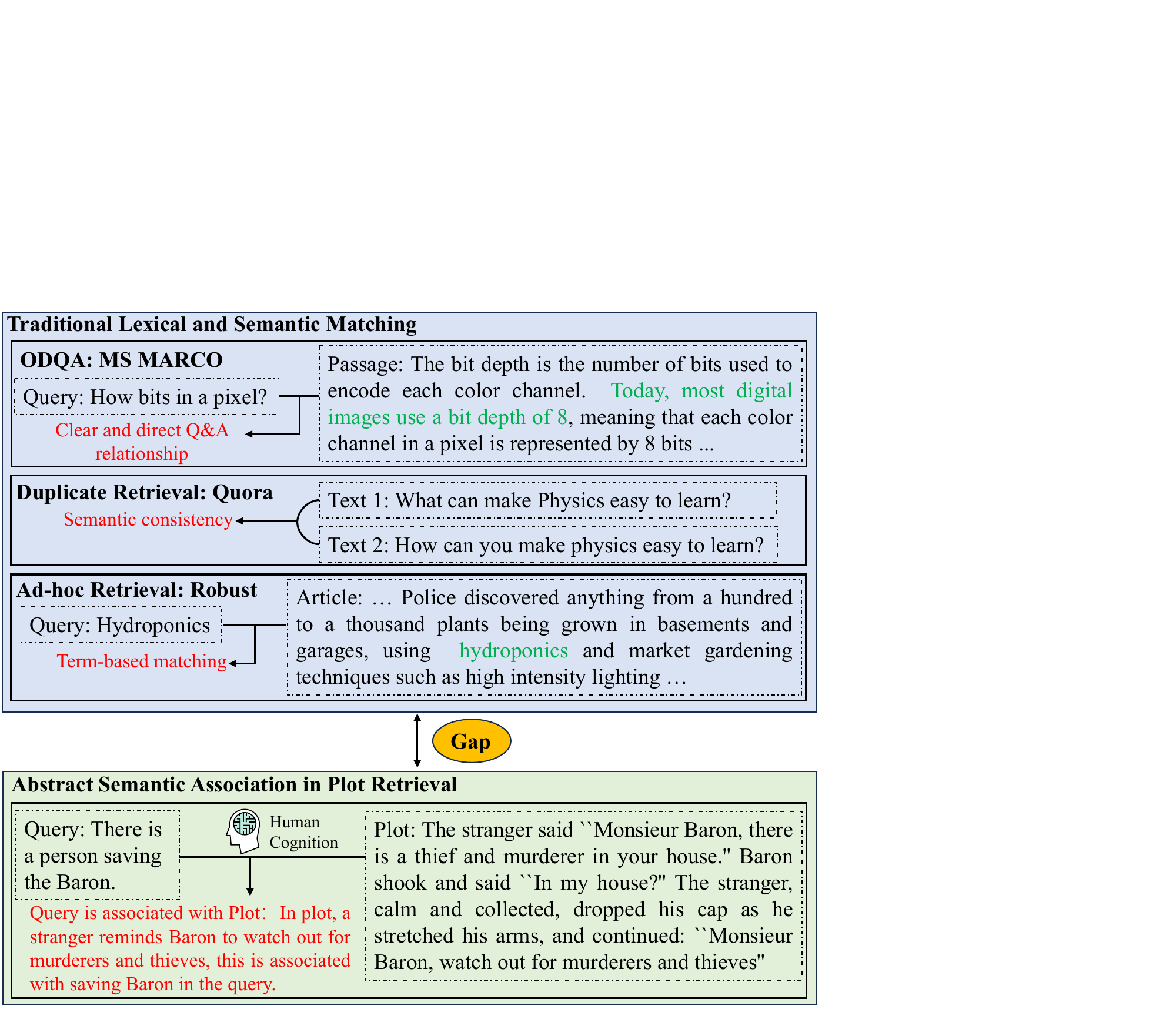} 
\caption{The gap between existing IR datasets and \taskname in estimating the relationship between texts.}
\label{comparision}
\end{figure}
Although \textit{Plot Retrieval} can be formalized as an information retrieval (IR) task, the key challenge in \taskname is estimating the abstract semantic association between two texts that cannot be simply measured by lexical or semantic matching. Specifically, we analyze the logs of online reading apps such as Kindle, iReader, Douban\footnote{\scriptsize{\url{https://www.ireader.com.cn}, \url{https://book.douban.com}.}} and find that the semantic association between the description of the plot given by the reader (i.e., query) and the actual plot in the book is very abstract. This abstract association is mainly because users integrate their own understanding, summaries, or speculations of the plot when writing the query, which makes it hard to directly associate plots to the query like traditional lexical matching, semantic similarity, or relevance. For example, for a plot: \textcolor{red!80!white}{\textit{The stranger said ``Monsieur Baron, there is a thief and murderer in your house.'' Baron shook and said ``In my house?'' The stranger, calm and collected, dropped his cap as he stretched his arms, and continued: ``Monsieur Baron, watch out for murderers and thieves''}}, and the query for this plot given by reader is \textcolor{blue!80!white}{\textit{There is a person saving the Baron.}} 
 % In the plot, \textcolor{red!80!white}{\textit{the stranger reminds Baron to watch out for murderers and thieves}}, this has the abstract semantic association with \textcolor{blue!80!white}{\textit{saving the Baron}} in the query.
 This association is generated by human cognition and is more difficult to estimate than just lexical or semantic matching because it requires IR models to understand that \textcolor{red!80!white}{\textit{the stranger reminds Baron to watch out for murderers and thieves}} is actually associated with \textcolor{blue!80!white}{\textit{saving the Baron}}, even though their literal meanings are different. However, as shown in Figure~\ref{comparision}, existing IR datasets do not reflect this abstract semantic association well. For example, in Open-domain Question-Answering such as MS MARCO~\cite{msmarco}, Natural Questions~\cite{nq}, and SQuAD~\cite{squad}, the query and its corresponding passage form a clear and direct question-and-answer relationship. In Duplicate Retrieval such as Quora and MRPC~\cite{MRPC}, the annotation is based on whether the semantics of the two texts are consistent. In Ad-hoc Retrieval such as Robust04~\cite{robust04}, lexical matching still accounts for the main part and semantic association is less~\cite{match-prompt}.
 
 A dataset that can reflect abstract semantic associations between texts generated by human cognition is important for the entire IR community to study the upper limit of the IR models' ability to model semantic association. However, it is very difficult to obtain the annotated query-passage pairs with sufficient abstract semantic association. Annotating abstract semantic association pairs requires annotators to pay the high reading cost for passage, and have sufficient comprehension ability to write a query that looks very different from the passage but has abstract semantic association with it.

In this paper, for \textit{Plot Retrieval}, a novel and challenging IR task, we propose a labeled dataset called \datasetname with 430K query-plot pairs. Compared with existing IR datasets, text pairs in \datasetname have the following obvious characteristics: (1) more abstract semantic association generated by human cognition and (2) less word overlap. These two characteristics enable \datasetname not only to be used to perform training on \textit{Plot Retrieval} task but also become the benchmark for evaluating the ability of IR models to estimate abstract semantic association between texts. In the construction of \datasetnamens, we collect publicly available raw data from the Internet, which shares the idea with~\cite{wan2019fine,personality}. To address the difficulty in annotation mentioned above, instead of directly asking the annotators to write a query that has abstract semantic association with the plot, we first use weakly supervised information to collect query-plot pairs that may have semantic association, and let the annotator select the pairs that really contain abstract semantic association, regularize these pairs, and get the final query-plot pairs.

In experiments, first, we evaluate various lexical retrieval, sparse retrieval, dense retrieval, and cross-encoder methods trained on mainstream IR datasets such as MS MARCO on \datasetnamens, and find that these methods do not perform well, which shows the difference between \datasetname and the current IR datasets. A noteworthy finding is that BM25, the strong zero-shot IR baseline based on lexical-matching~\cite{beir,contriever}, achieves better performance on BEIR~\cite{beir} than many neural IR models, but has worse performance on \datasetnamens. This indicates that \datasetname has the higher challenge for semantic understanding rather than simple literal matching. Second, we train IR models on our weakly supervised data and achieve better performance than the models trained on MS MARCO, which indicates the effectiveness of our annotation strategy. Third, human studies show that the current IR models are far behind human in capturing abstract semantic association, and there is a lot of room for improvement in future research. Our contributions are:

\noindent$\bullet$ We propose a novel, critical and challenging task called \textit{Plot Retrieval}, design a novel evaluation metric called N-RODCG and construct a dataset called \datasetname for this task. 

\noindent$\bullet$ Extensive experiments across various IR models and the comparison with human studies on \datasetname show that the current IR models still struggle in capturing abstract semantic association between texts and there is a lot of room for improvement in the future research.

\noindent$\bullet$ We broaden the research field of Information Retrieval from lexical or semantic matching to more ambiguous abstract semantic association between texts, and \datasetname can be used as an effective benchmark for evaluating this ability of IR models. We will release both English and Chinese versions of \datasetname at \url{https://github.com/xsc1234/Plot-Retrieval} for further research.
% \begin{itemize}
% \item{} We propose a novel, critical and challenging task called \textit{Plot Retrieval} and construct a dataset called \datasetname for this task. 
% \item{} Extensive experiments across various lexical retrieval, sparse retrieval, dense retrieval and cross-encoder methods compared with human studies on \datasetname show that the current state-of-the-art IR models still struggle in capturing abstract semantic association between texts and there is a lot of room for improvement in the future research.
% \item{} We broaden the research filed of Information Retrieval from lexical or semantic matching to more ambiguous abstract semantic association between texts, and \datasetname can be used as an effective benchmark for evaluating this ability of IR models. 
% \end{itemize}

% 为了完成plotretrieval，一个对IR模型抽象语义关联能力有较大挑战性的任务，我们收集了数据集

% 我们将信息检索社区的关注点从简单的词或语义匹配拓宽到更加模糊的抽象语义关联，为进一步探索神经信息检索模型对文本间抽线关联能力建模的研究提供了重要的测试基准

\section{Related Work}

\paragraph{Information Retrieval Datasets}
According to specific task, existing mainstream IR datasets can be divided into: \emph{Open Domain Question-answering} (MS MARCO~\cite{msmarco}, Natural Questions~\cite{nq}, TriviaQA~\cite{triviaqa}, SQuAD~\cite{squad}, WebQuestions~\cite{webq}, FiQA~\cite{fiqa}, HotPotQA~\cite{hotpotqa} and CuratedTREC~\cite{trec}, etc.), \emph{Ad-hoc Retrieval} (Robust~\cite{robust04}, ClueWeb~\cite{clueweb}, MQ2007~\cite{mq2008}), \emph{Duplicate Retrieval} (Quora, CQADupStack~\cite{cqadupstack}, MRPC~\cite{MRPC}), \emph{Entity Retrieval} (DBPedia-Entity~\cite{dbpedia}), \emph{Argument Retrieval} (ArguAna~\cite{arguana} and Touchè-2020~\cite{touche}), \emph{Citation Prediction} (SCIDOC~\cite{scidocs}) and \emph{Fact Checking} (FEVER~\cite{fever} and Climate-FEVER~\cite{climatefever}).
Existing datasets also cover a range of different domains of target documents like \emph{Bio-Medical articles}~\cite{biosq},
\emph{Tweets}~\cite{signal}, \emph{News}~\cite{trec-news}.

In all the above datasets, the matching between texts can be summarized as a combination of lexical and semantic matching. The relationship of query-passage pairs in these datasets can usually be judged only by the literal meaning, without the need to deeply understand the semantics and judge the abstract association between semantics. Direct evidence is that BM25~\cite{bm25} can significantly defeat many neural IR models that have been trained on large-scale supervised datasets only through lexical matching on these datasets in the zero-shot setting~\cite{beir}. \datasetname has more abstract semantic association and less word overlap between texts, which is a more challenging dataset for IR models.

\paragraph{IR Datasets for Books}
Our dataset also extends into the significant domain of narrative literature for IR applications. While there exists an extensive list of datasets on story understanding (for more details, please refer to the survey~\cite{sang2022survey}), there has been limited work addressing the IR aspect within the context of stories.
In relation to our work, two other datasets are noteworthy. The first is \emph{RELiC}~\cite{relic}, which frames the task as utilizing literary analysis paragraphs to retrieve quoted text. This task essentially falls within the realm of IR, although it lacks a standard format of IR queries.
The second is \emph{NarrativeQA}~\cite{kovcisky2018narrativeqa}, primarily designed as a book QA dataset but adaptable for an IR task~\cite{frermann2019extractive,mou2021narrative}. However, it comes with a limitation that it does not provide groundtruth for the retrieval purposes.

% \noindent \textbf{Datasets for book Reading. }

\section{Task Description}
\subsection{Abstract Semantic Association}

\begin{figure}[t]
\centering
\includegraphics[width=\linewidth]{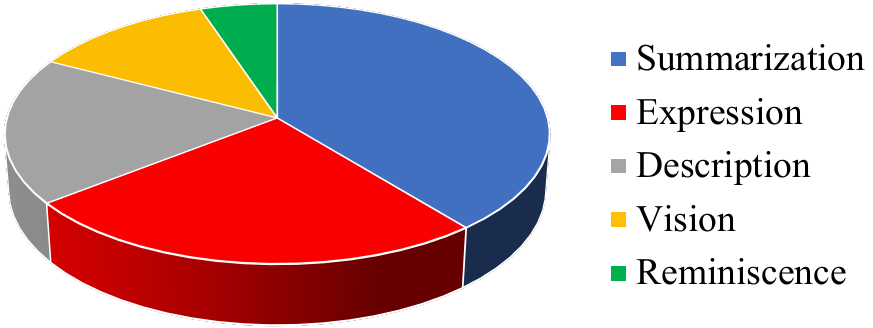} 
\caption{Statistics of abstract semantic association.}
\label{bingtu}
\end{figure}

In the analysis of public data of online reading apps, we conclude five main manifestations of abstract semantic association between the query and the plot. (1) Query abstractly summarizes the plot (Summarization). (2) Query expresses feelings, analysis or comments about the characters or events in the plot (Expression). (3) Query depicts the characters in the plot (Description). (4) Query describes the overall visual information formed by the environment, characters, and events in the plot (Vision). (5) Query is motivated by the event in the plot to reminisce another related event (Reminiscence). Their statistics are shown in Figure~\ref{bingtu}.

\subsection{Task Definition}
\textit{Plot Retrieval} aims to retrieve the relevant plots from the book for a query. Specifically, given a query $q$, a candidate set of plots $P = \{p_1,p_2,...,p_n\}$ for a book and each plot $p_i$ consists of $m$ sentences ($m$ is a hyperparameter and we set it as $3$). The model needs to give the ranking score for each $p_i \in P$ based on the association between plot $p_i$ and query $q$, rank the plots in $P$ according to the score, and return a list $R$ with Top-K plots. The challenge of this task is mainly in two aspects: (1) The semantic association between the query and the plot is very abstract. This is mainly because users integrate their own understanding, summaries, or speculations of the plot when writing descriptions. IR models struggle in identifying this abstract association. (2) Plots in the candidate set $P$ come from the same book, they have semantic and entity relatedness to each other. It makes IR models hard to distinguish the semantic difference.
% \mo{concretize how to set the length of $p_i$ and how to split the book into $p_i$s}

\subsection{Evaluation Metric: N-RODCG}
As for the evaluation metrics for \tasknamens, in addition to the common information retrieval metrics, such as MRR (Mean Reciprocal Rank) and Recall, we propose N-RODCG (Normalized Relative Offset Distance Discounted Cumulative Gain), a novel metric that is more in line with the actual reading scene. The motivation of this metric is that each plot of the candidate set is actually the segment of continuous texts in the original book, even if the retrieved plot is not exactly the ground-truth plot, as long as it is close enough to the ground-truth plot in the original book, the ground-truth plot will appear in the reader's field of vision and be noticed by the reader. In addition, there is the strong semantic association between plots with small distances. N-RODCG comprehensively measures the ranking of the plots in $R$ and their distance from the ground-truth plots. For a query $q$, given a retrieved list of plots $R = [p_1',p_2',...,p_k']$ obtained from the model. Because each plot $p_i'$ consists of $m$ sentences, we can get the position of $p_i'$ in the original text of the book, which is the average value of each sentence index in $p_i'$ and we call it $s_i$. Then the positions of the plots in $R$ are $S=[s_1,s_2,...,s_k]$. And the positions ($t_i$) of the ground-truth plots for $q$ is $T=[t_1,t_2,...,t_g]$, $g$ is the number of ground-truth plots. The relative offset distance $d_i$ between $p_i'$ and ground-truth plots of $q$ can be computed as:
\begin{equation}
    d_i = \min(|s_i-t_1|,|s_i-t_2|,...,|s_i-t_g|).
\end{equation}
Then, we define the Discounted Cumulative Gain~\cite{dcg} between ROD and the ranking of the retrieved plots:
\begin{equation}
    \textrm{RODCG}@k = \sum_{i=1}^{k} {f(d_i) \over \log(i+1)}, 
\end{equation}
where $i$ is the ranking of plot $p_i'$, $f$ is the piecewise function ($\alpha$ is the window of the reader’s field of vision and we set it to $5$ based on statistical data):
\begin{equation}
    f(d_i) =
\begin{cases} 
{1 \over {d_i+1}},  & d_i < \alpha ; \\
0, & otherwise .
\end{cases}
\end{equation}
N-RODCG can be computed as:
\begin{equation}
    \textrm{N-RODCG}@k = {\textrm{RODCG}@k \over \textrm{I-RODCG}@k}, 
\end{equation}
I-RODCG is the value when the plots in retrieval list $R$ for $q$ are optimally ranked, that is, the theoretical maximum value of N-RODCG.

\section{\datasetname}
We introduce collection, filtering, translation, annotation, and statistics for \datasetname in this section. More details are introduced in Appendix.

\subsection{Overview of Dataset Construction}
The row data of \datasetname is collected from an online reading app on the Internet. Specifically, we notice recent reading apps allow readers to write publicly available comments on the texts in the book. Many of these comments include abstract descriptions of the plots in the corresponding texts. They are written by the readers based on their own understanding during book reading. While they are semantically associated with the plots, they require sufficient comprehension ability to discover and are challenging for IR models to identify. These comment-plot pairs constitute the weakly supervised signal for query-plot pairs in \datasetnamens. We first filter these pairs to remove the comments that have obvious word overlap with plots or have little practical meaning. However, the filtered comment-plot pairs still cannot be directly used as \datasetnamens, because the comments written by readers are free-style and have a lot of noise. We let the annotators do more identification and rewriting on them. After the human annotation, we exploit the labeled datasets to construct an automatic annotation model for fast, low-cost acquisition of large datasets. Last but not least, we ensure the complete independence of the training set and the test set during the construction of \datasetnamens, which makes that there are enough differences in the domain between the training set and the test set to more reasonably evaluate the ability of the IR models to estimate abstract semantic association.

\subsection{Dataset Construction} \label{construction}

\hypertarget{collection}{\noindent \textbf{Step 1: Data Collection. }We collect data for training set and test set separately. Specifically, for test set, we use 33 publicly available English books that are collected from Gutenberg project and processed by ~\cite{personality}. We find 84 Chinese versions of these 33 English books that we have licenses of usage. We sample 52,924 public comments written by readers for various plots in these 84 books. For the training set, we collect 105 books from the same reading app and sample 1,005,480 comments. There is no overlap between books in the training set and the test set. }

\hypertarget{filtering}{\noindent \textbf{Step 2: Data Filtering. }Before human annotation, we perform a preliminary filter on the collected data. Specifically, first, in order to make the description of the comment for the plot abstract enough, we remove the comments that have a lot of word overlap with the original texts in the book. Given a comment $c$ and the original text $t$ in the book marked by the comment $c$, we use NLTK\footnote{https://www.nltk.org/} to perform word tokenization on them and remove the stop words. Then we get the sets of words for them ($\mathbb{C}$ and $\mathbb{T}$). We remove the comments that:
\begin{equation}
   {|\mathbb{C} \cap \mathbb{T}| \over |\mathbb{C}|}  > 0.5.
\label{over}
\end{equation}
Second, we remove the comments that have little practical meaning. That is, the comments that do not describe the plot but express the reader's emotions such as ``\emph{This is so funny!}'' or ``\emph{I can't understand this}''. We use ChatGPT\footnote{https://openai.com/blog/chatgpt} via prompting it to judge whether the comment is describing a specific plot rather than simply expressing emotion to complete this task. Considering that a large amount of data will bring high ChatGPT usage cost, we perform this filtering operation on the full test set and 50,000 samples of training set. For the other samples in the training set, we use the automatic annotation model for fast and low-cost filtering, which will be introduced in \hyperlink{aam}{Step 5}. After this, we get 7,661 samples in test set and 7,432 samples in training set for human annotation.}

\hypertarget{annotation}{\noindent \textbf{Step 3: Human Annotation. }For the sample with a comment $c$ and the original text $t$ in the book marked by the comment $c$, annotators have two tasks to finish. (1) Judge whether $c$ contains the abstract description of the plot in $t$. (2) If so, mark the texts describing the plot from $c$ and use the texts as the query $q$. After this step, we can get the query-plot pairs where there is the abstract semantic association between query and plot. } Specifically, we first select nine annotators who have at least a high school education level, because our task requires the annotators to have a certain ability to understand literary works. We write the guidelines to help the annotators better understand the details of the annotation task. Before the formal annotation start, we conduct three rounds of pre-annotation and verify the pass rate of each annotator's work. We select the annotator whose pass rate of work reaches 90\% in the pre-annotation for formal annotation. In the formal annotation, for the results of each annotator, we introduce another annotator to sample and validate the results and give the pass rate, which can measure whether two annotators agree with the results. We continue to screen and guide the annotators until the pass rate of each annotator reaches 95\%. We select the samples that $c$ are judged to contain abstract descriptions of $t$ as the final samples. After this, we get 4,572 query-plot samples in the test set and 4,402 samples in the training set.

\hypertarget{trans_and_corpus}{\noindent \textbf{Step 4: Translation and Corpus Construction. }Since the majority of our collected data is in Chinese, we translate the collected data into English. For test set, all books have their public English versions (\hyperlink{collection}{Step 1}). So we (1) translate the comment $c$ to English and (2) project the original text $t$ in the Chinese book marked by the comment $c$ to its content in the English version of the book. For the first task, we finish it by ChatGPT. For the second task, we use Spacy to sentencize the texts of books, use multilingual embedding LASER\footnote{https://github.com/facebookresearch/LASER.} to embed sentences and use \textit{vecalign}~\cite{vecalign} to align the sentences between books based on sentence embeddings. For training set, because some books do not have the corresponding English versions, we directly translate $c$ and $t$ to English by \textit{Helsinki}\footnote{https://huggingface.co/Helsinki-NLP}, a neural machine translation model.}

We use the collection of plots of books in the test set as the retrieval corpus, which means that when we test the retrieval performance of the IR models on \datasetnamens, the samples in the training set do not appear in any test data. For the book, we divide every $m$ sentences into a chunk (the basic unit of the corpus). We mark the chunks containing the sentences in $t$ as ground truth for $c$. To ensure the semantic integrity of $t$, we also make $t$ as a chunk and mark it as ground truth. Details of the corpus are shown in Appendix~\ref{corpus_detail}.

\begin{table}[t]
\setlength\tabcolsep{10
pt}
\centering
\scalebox{0.85}{
\begin{tabular}{lr}
\toprule
\#Train Pairs    & 400,000 \\
\#Validation Pairs & 37,609  \\
\#Test Quries    & 4,572   \\
\#Candidate plot chunks  & 136,195 \\ \hline
Average query length & 29.12 \\
Average chunk length  & 58.10 \\
\toprule
\end{tabular}
}
\caption{Statistics of \datasetname.}
\label{statistics}
\end{table}

\begin{table}[t]
\setlength\tabcolsep{10
pt}
\centering
\scalebox{0.85}{
\begin{tabular}{lc}
\toprule
Dataset        & Word Overlap \\ \hline
FEVER          & 61.57        \\
Quora          & 53.75        \\
Touché-2020    & 51.77        \\
SCIFACT        & 48.24        \\
MS MARCO       & 46.29        \\
Dbpedia        & 41.54        \\
FiQA-2018      & 38.40        \\
NQ             & 36.24        \\
HotPotQA       & 35.66        \\
Climate-Fever  & 29.02        \\
Arguana        & 28.98        \\
SCIDOCS        & 26.79        \\
Trec Covid     & 26.41        \\
%REiLC     & 24.21        \\
NFCorpus       & 23.33        \\
\datasetnamens & \textbf{19.62}    \\
\toprule
\end{tabular}}
\caption{Word overlap between query and the positive candidate documents among various IR datasets.}
\label{word overlap}
\end{table}

\hypertarget{aam}{\noindent \textbf{Step 5: Auto Annotation Model. }For the large amount of data in the training set that has not been manually annotated, we construct a text-pair binary classifier to complete automatic annotation. Specifically, we train BERT\footnote{https://huggingface.co/bert-base-uncased}~\cite{bert} on 50,000 samples of training set in \hyperlink{filtering}{Step 2} in which 4,402 are annotated as positives in \hyperlink{annotation}{Step 3} and the other are negatives. We use the trained classifier to automatically annotate the data in the training set. Although most of the data in the training set is constructed under the weak supervision of the automatic annotation model, experiments in Section~\ref{exp_res} show that compared with large-scale supervised IR datasets, our training data is better for IR models to estimate the abstract semantic association.}

\subsection{Data Statistics}
Table~\ref{statistics} shows the statistics of the training set and test set in \datasetnamens. Most of the train and validation pairs are obtained from the auto annotation model in \hyperlink{aam}{Step 5}. Table~\ref{word overlap} shows the word overlap between the query and candidate documents (calculated by Equ (\ref{over})). \datasetname has the lowest overlap, especially compared to mainstream IR datasets such as MS MARCO. Therefore, compared to the existing IR datasets, the query-plot pairs in \datasetname pose a higher challenge to the IR models. The pairs look very different but have abstract semantic association, rather than simple lexical or semantic matching.

\section{Experiments}
\label{experiment}
In this section, we evaluate various IR models on \datasetname and perform human studies.

\subsection{Baselines}
%Details of baselines are introduced in Appendix.

\noindent \textbf{Lexical Retrieval. } We use (1) \textbf{BM25}~\cite{bm25}, a a bag-of-words retrieval method based on word-to-word exact matching.

\noindent \textbf{Sparse Retrieval.} Following BEIR~\cite{beir}, we select three mainstream sparse retrieval models including (1) \textbf{DeepCT} (learning dynamic term weights)~\cite{deepct}, (2) \textbf{SPARTA} (learning a sparse representation that can be efficiently implemented as an inverted index)~\cite{sparta} and (3) \textbf{DocT5query} (generating queries added to documents)~\cite{doct5query}. All of them are fine-tuned on MS MARCO.

\noindent \textbf{Dense Retrieval.} (1) \textbf{DPR}~\cite{dpr}, a classical dense retrieval model based on bi-encoder and trained with BM25 hard negatives and in-batch contrastive loss. (2) \textbf{ANCE}~\cite{ance}, it dynamically updates negatives during training. (3) \textbf{TAS-B}~\cite{tas-b} is trained with supervision from cross-encoder. (4) \textbf{BERM}~\cite{berm,xuberm}, a plug-and-play method to enable dense retrieval models to learn representations that are more suitable for matching. (5) \textbf{Ernie-Search}~\cite{erniesearch} trains dense retrieval model by cascade distillation from ColBERT~\cite{colbert} and cross-encoder. All of the above baselines are fine-tuned on MS MARCO. There are also some methods first pre-train models on large-scale datasets by self-supervised IR signal. (6) \textbf{COCO-DR}~\cite{coco} is pre-trained on BEIR~\cite{beir}. (7) \textbf{coCondenser}~\cite{cocondenser} and (8)  \textbf{RetroMAE}~\cite{retromae} are pre-trained on English Wikipedia and BookCorpus. (8) \textbf{Contriever}~\cite{contriever} is pre-trained on English Wikipedia and CCNet. All of these models are fine-tuned on MS MARCO after pre-training for IR.

\noindent \textbf{Late-Interaction.} \textbf{ColBERT}~\cite{colbert} performs late interaction on embeddings of each token to achieve finer-grained interaction than dense retrieval. This model is fine-tuned on MS MARCO.

\noindent \textbf{Re-Ranking.} We use \textbf{Cross-Encoder}~\cite{cross-encoder} that exploits self-attention for interaction between tokens as re-ranker, which has shown power in Book QA tasks~\cite{mou2021narrative}. Before re-ranking, we first use Contriever to retrieve Top-100 documents for each query as its candidate list. This model is fine-tuned on MS MARCO.

\noindent \textbf{ChatGPT-Assisted.} ChatGPT performs well on various NLP tasks, we also explore its performance on \tasknamens. It is expensive to directly let ChatGPT inference on a large-scale corpus, so we prompt ChatGPT to generate the plot in the corresponding book for the query (query expansion~\cite{query_expan}), and then use the generated plot as query and use Contriever to retrieve related plots from the corpus.

\subsection{Experimental Settings}
\begin{table*}[t]
\renewcommand\arraystretch{1}
%% 设置表格每一行
\setlength\tabcolsep{9
pt}%调列距
%% 设置单元格内行间距
\centering
\scalebox{0.9}{
\begin{tabular}{llllllllll}
\toprule
Model         & \multicolumn{3}{c}{MRR} & \multicolumn{3}{c}{Recall} & \multicolumn{3}{c}{N-RODCG} \\ 
              & @1   & @10     & @100    & @1   & @10    & @100   & @1     & @10     & @100     \\ \hline
\multicolumn{10}{c}{Lexical Retrieval}                                                                                     \\ 
BM25           &0.063    & 0.093        &  0.100       & 0.063     & 0.083       &  0.182       & 0.077     & 0.085        & 0.125         \\ \hline
\multicolumn{10}{c}{Sparse Retrieval}                                                                                      \\ 
SPARTA        & 0.059    & 0.090         & 0.098        & 0.059   & 0.096   & 0.253       & 0.069       & 0.088    &  0.143               \\
DeepCT        & 0.043    & 0.085         & 0.091        & 0.043   & 0.089   & 0.242       & 0.058       & 0.082    &  0.136             \\
docT5query     & 0.085  & 0.124        & 0.136        & 0.085  & 0.130    & 0.330       & 0.107       & 0.129     & 0.199             \\ \hline
\multicolumn{10}{c}{Dense Retrieval}                                                                                       \\ 
DPR           & 0.081   & 0.123        &  0.132       & 0.081   & 0.129   & 0.321       & 0.098       & 0.121    & 0.193            \\
ANCE           & 0.088  & 0.129        & 0.139        & 0.088  & 0.136    & 0.332       & 0.110       & 0.132     & 0.204             \\
TAS-B$^{\bullet}$         & 0.091     & 0.140        & 0.150       & 0.091  & 0.161    & 0.373        & 0.112        & 0.148    & 0.227           \\
BERM          & 0.088       & 0.132       & 0.141        & 0.088   & 0.149   & 0.354       &  0.107       & 0.137   & 0.214             \\
coCondenser$^{\star}$  &  0.097       & 0.146        & 0.155        & 0.097   & 0.162   & 0.368       & 0.116       &  0.151    & 0.227             \\
Ernie-Search$^{\bullet}$  &  0.102       & 0.151        & 0.161        & 0.102   & 0.167   & 0.381       & 0.124       & 0.158    & 0.238          \\
Contriever$^{\star}$    &  0.111       & 0.165        & 0.175        & 0.111   & \underline{0.184}   & \bf 0.416       & 0.137       & \underline{0.176}    & \underline{0.262}            \\
COCO-DR$^{\star}$       &  0.096      & 0.145        & 0.155        & 0.096  &  0.158   &  0.375      & 0.118       & 0.150    &  0.231          \\
RetroMAE$^{\bullet \star}$      &  0.108       & 0.158        &  0.168        & 0.108   & 0.174    & 0.395        & 0.132       & 0.168     & 0.249           \\ \hline
\multicolumn{10}{c}{Late-Interaction}                                                                                      \\ 
ColBERTv2       & \underline{0.120}        &  \underline{0.170}      & \underline{0.179}        & \underline{0.120}  & 0.144    & 0.290       &  \underline{0.141}      & 0.151    & 0.211           \\ \hline
\multicolumn{10}{c}{Re-Ranking}                                                                                            \\ 
Cross-Encoder & \textbf{0.123}        & \textbf{0.174}        & \textbf{0.184}        & \textbf{0.123}   & \textbf{0.197}    & \textbf{0.416}       & \textbf{0.150}       & \textbf{0.189}   & \textbf{0.272}            \\ \hline
\multicolumn{10}{c}{ChatGPT-Assisted}                                                                                            \\ 
ChatGPT+Contriever &0.048         & 0.077        & 0.085        & 0.048  & 0.088    &  0.254      & 0.062       &  0.083  & 0.142         \\ \toprule
\end{tabular}
}
\caption{Zero-shot performance of IR models on test set of \datasetnamens. \textbf{Bold}: best performance. \underline{Underlined}: second best performance. $\star$: Train on large self-supervised data. ${\bullet}$: Knowledge distillation from cross-encoder.}
\label{zero-shot}
\end{table*}
\textbf{First}, to explore the ability of the SOTA IR models trained on MS MARCO to estimate abstract semantic associations between texts, we evaluate the performance of them in zero-shot setting on the English version of \datasetnamens. \textbf{Second}, to show the effectiveness of our weakly supervised training data, we compare the performance of IR models trained on weakly supervised training data in \datasetname with existing IR datasets in the same training method and settings. We use \textit{bert-base-uncased} and \textit{bert-base-chinese} as pre-trained models for English and Chinese respectively. In training, we set the learning rate to $10^{-5}$. We train the model with $64$ batch size on a single A100 GPU for 5 epochs and use Pytorch~\cite{pytorch} as the training framework. \textbf{Third}, the difficulty of \datasetname for IR models can be reflected by the performance gap between IR models and humans on different datasets. We compare this gap on different IR datasets via human studies. 

\subsection{Experimental Results} \label{exp_res}

\begin{table*}[t]
\renewcommand\arraystretch{1}
%% 设置表格每一行
\setlength\tabcolsep{5
pt}%调列距
%% 设置单元格内行间距
\centering
\scalebox{0.9}{
\begin{tabular}{lllllllllll}
\toprule
Dataset & Domain & \multicolumn{3}{c}{MRR} & \multicolumn{3}{c}{Recall} & \multicolumn{3}{c}{N-RODCG} \\ 
        &       & @1  & @10     & @100    & @1   & @10    & @100   & @1  & @10     & @100     \\ \hline
\multicolumn{11}{c}{English Setting}                                                                                     \\ 
MS MARCO  & Misc.      & 0.080   & 0.121        &  0.131       & 0.080   & 0.125   & 0.320       & 0.095       & 0.119    & 0.190            \\
RELiC     & Book      & 0.083   & 0.128    &  0.138       &  0.083       & 0.142  & 0.389    & 0.102       &  0.134      & 0.225           \\ 
\makecell[l]{\datasetname \\ (weakly supervised)} & Book & \textbf{0.105$^{\dag}$}   & \textbf{0.155$^{\dag}$}    &  \textbf{0.165$^{\dag}$}       & \textbf{0.105$^{\dag}$}        & \textbf{0.174$^{\dag}$}   & \textbf{0.420$^{\dag}$}    & \textbf{0.128$^{\dag}$}       & \textbf{0.163$^{\dag}$}       &  \textbf{0.253$^{\dag}$}       \\ \hline
\multicolumn{11}{c}{Chinese Setting}                                                                                      \\ 
DuReader    & Misc.    & 0.031  & 0.041     & 0.045        & 0.031        & 0.062   & 0.175   & 0.041       & 0.075    & 0.139           \\
\makecell[l]{\datasetname \\ (weakly supervised)}  & Book     & \textbf{0.103$^{\dag}$}  &  \textbf{0.152$^{\dag}$}    &  \textbf{0.164$^{\dag}$}       & \textbf{0.103$^{\dag}$}        & \textbf{0.247$^{\dag}$}   &  \textbf{0.588$^{\dag}$}       & \textbf{0.140$^{\dag}$}       & \textbf{0.169$^{\dag}$}     & \textbf{0.257$^{\dag}$} \\\toprule
\end{tabular}
}
\caption{Performance of the (DPR) models trained on different IR datasets on test set of \datasetnamens. \textbf{Bold}: best performance. ${\dag}$: significant performance improvement with p-value $ \leq 0.05$ compared with baselines.}
\label{compare_dataset}
\end{table*}

\noindent \textbf{Performance on \datasetnamens.} Table~\ref{zero-shot} shows the zero-shot performance of IR models trained on MS MARCO on test set of \datasetnamens. We can draw the following four conclusions. \textbf{(1)} \datasetname has more abstract semantic association and less word overlap between texts than existing IR datasets, which is more challenging for current SOTA IR models. This can be supported by the phenomenon that BM25, the strong zero-shot IR baseline based on term-matching~\cite{beir,contriever}, achieves better performance on BEIR~\cite{beir} than many neural IR models such as DPR, ANCE, and TAS-B, but has worse performance on \datasetname than all neural IR baselines that can capture the semantic matching information. \textbf{(2)} More training data facilitates the estimation of abstract semantic association, even if the data is self-supervised. This can be supported by the phenomenon that models pre-trained on large-scale datasets such as coCondenser, Contriever, COCO-DR, and RetroMAE have better performance than the models fine-tuned directly on MS MARCO. \textbf{(3)} More interactions between texts are conducive to the estimation of abstract semantic association. Cross-Encoder that exploits self-attention for fine-grained interaction between tokens shows the best performance. \textbf{(4)} ChatGPT is not good at associating plots with their abstract corresponding queries. Using ChatGPT to generate the plot associated with the query, and using the generated content as the new query for retrieval by Contriever achieves worse performance. It is because we find that ChatGPT cannot accurately generate the plots associated with the query but generates the common content for the book such as the summary and background of the book. This makes the query ambiguous and indiscriminate.

\noindent \textbf{Discussion on N-RODCG.} In this paper, we propose a new evaluation metric named N-RODCG, which is more in line with the actual book reading scene. Specifically, traditional IR metrics such as MRR, Recall and NDCG can only reflect the difference in relevance between the texts in the returned rankted list and the ground-truth. However, in the book reading scene, a more reasonable metric is to reflect the distance between the retrieved texts and the ground-truth in the book. Because this can better reflect the retrieval models' ability to help readers find the content they want from the book. The greater the value of the metric, the closer the retrieved texts is to the ground-truth in the book, and the easier for readers to find what they want to read. 

\noindent \textbf{Effect of Weakly Supervised Training Data.} The weakly supervised training data we construct has positive significance for improving the performance of the IR models on the task \tasknamens. Specifically, we compare the performance of models trained on mainstream supervised datasets (human annotation) with the models trained on weakly supervised training data in \datasetnamens. In English setting, we use two datasets as baselines. The one is MS MARCO, the large-scale labeled IR dataset. The other is RELiC~\cite{relic}, the large-scale labeled IR dataset that aims to retrieve evidence for literary claims, whose domain also involves book reading. In Chinese setting, we use DuReader~\cite{dureader}, a large-scale Chinese labeled IR dataset. These models are fine-tuned with the same method (DPR) and settings and perform early stopping on validation pairs. Table~\ref{compare_dataset} shows that weakly supervised training data in \datasetname significantly improves the performance of the IR models on \taskname than mainstream supervised IR datasets with much more human annotations. We maintain the independence of the training set and test set in the process of data construction so that there is enough domain gap between them. Besides, although RELiC also belongs to the book domain, its performance is not significantly improved compared with MS MARCO. This further shows the effectiveness of our weakly supervised training data for IR models to learn the abstract semantic association between texts instead of just overfitting the domain.

\begin{figure}[t]
\centering
\includegraphics[width=\linewidth]{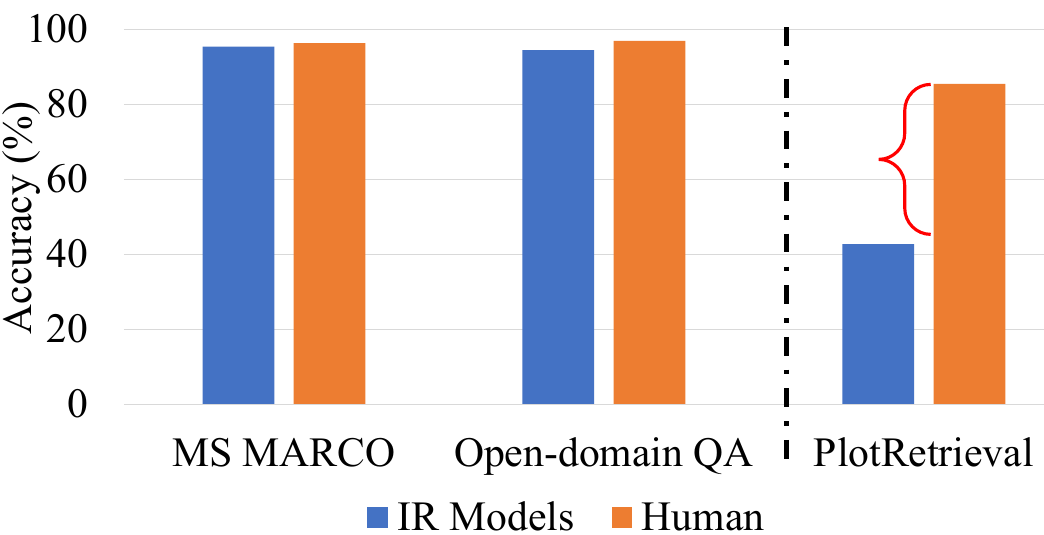} 
\caption{The gap between IR models and humans.}
\label{human}
\end{figure}
\noindent \textbf{Human Studies. }We perform human studies to compare the performance gap of IR models and humans on MS MARCO, ODQA (consisting of Natural Questions, TriviaQA, SQuAD, WebQuestions), and \datasetnamens. Specifically, we sample $500$ queries from the test sets of these three datasets respectively, for each query, we construct a candidate list containing 1 ground truth and 19 negatives. We let the IR model and humans select the ground truth for the query from its candidate list and count the accuracy. We use DPR (trained on MS MARCO) for MS MARCO, DPR (trained on ODQA) for ODQA, and Cross-Encoder (the best model in Table~\ref{zero-shot} and trained on MS MARCO and \datasetname) for \datasetname as the IR models. We select three humans with college degrees for this study and count the average accuracy. Results in Figure~\ref{human} show that although the performance of the IR models on MS MARCO and ODQA is close to human, they still struggle in capturing abstract semantic association on \datasetnamens. 

\section{Conclusion}
In this paper, we propose a novel task called \taskname that retrieves relevant plots from the book for a query. Compared with the existing IR datasets, \taskname requires the IR models to have the strong ability to capture the abstract semantic association between texts rather than the simple lexical and semantic matching. It is meanly because readers integrate their own understanding, summaries, or speculations of the plot when writing the query. For the \taskname task, we propose \datasetnamens, a large labeled dataset with more abstract semantic association and less word overlap between texts, which can be used as a benchmark to train and evaluate the ability of IR models to capture abstract semantic associations between texts. Extensive experiments across various lexical retrieval, sparse retrieval, dense retrieval, and cross-encoder methods compared with human studies on \datasetname show that the current IR models still struggle in capturing abstract semantic association between texts and there is a lot of room for improvement in future research.

% \newpage
\section*{Limitations}
In this paper, we propose a novel task called \tasknamens. \taskname aims to retrieve the relevant plots for the query and has higher requirement for the ability of the information retrieval models to estimate the abstract semantic association between texts while existing information retrieval datasets are not satisfied. To achieve it, we collect and release \datasetnamens, a large-scale information retrieval dataset with more abstract semantic association and less word overlap. However, although comparison with humans shows that current SOTA IR models cannot perform well at this task, we do not propose an efficient solution such as novel model architecture and training method to solve this problem. Our contributions focus on proposing a more challenging retrieval task and dataset. Further research on the task will be carried out in future work.

\section*{Ethics Statement}
In the construction of datasets, we prioritize the ethical use of data and are committed to upholding the highest standards when it comes to protecting user privacy and ensuring data integrity. Specifically, all the data within our dataset is collected exclusively from publicly available information from online applications (apps). We strictly adhere to the legal guidelines and terms of service of these apps during the data collection process. Our data collection practices prioritize user privacy. All personally identifiable information (PII) has been thoroughly masked or removed from the dataset.  We declare that our work complies with the \href{https://www.aclweb.org/portal/content/acl-code-ethics}{ACL Ethics Policy}.

% \section*{Acknowledgements}
% This document has been adapted by Yue Zhang, Ryan Cotterell and Lea Frermann from the style files used for earlier ACL and NAACL proceedings, including those for 
% ACL 2020 by Steven Bethard, Ryan Cotterell and Rui Yan,
% ACL 2019 by Douwe Kiela and Ivan Vuli\'{c},
% NAACL 2019 by Stephanie Lukin and Alla Roskovskaya, 
% ACL 2018 by Shay Cohen, Kevin Gimpel, and Wei Lu, 
% NAACL 2018 by Margaret Mitchell and Stephanie Lukin,
% Bib\TeX{} suggestions for (NA)ACL 2017/2018 from Jason Eisner,
% ACL 2017 by Dan Gildea and Min-Yen Kan, NAACL 2017 by Margaret Mitchell, 
% ACL 2012 by Maggie Li and Michael White, 
% ACL 2010 by Jing-Shin Chang and Philipp Koehn, 
% ACL 2008 by Johanna D. Moore, Simone Teufel, James Allan, and Sadaoki Furui, 
% ACL 2005 by Hwee Tou Ng and Kemal Oflazer, 
% ACL 2002 by Eugene Charniak and Dekang Lin, 
% and earlier ACL and EACL formats written by several people, including
% John Chen, Henry S. Thompson and Donald Walker.
% Additional elements were taken from the formatting instructions of the \emph{International Joint Conference on Artificial Intelligence} and the \emph{Conference on Computer Vision and Pattern Recognition}.

\bibliography{anthology,custom}
\bibliographystyle{acl_natbib}
\newpage
\appendix

\section{Interface for Annotation}
\label{sec:appendix}
\begin{figure*}[h]
\centering
\includegraphics[width=\linewidth]{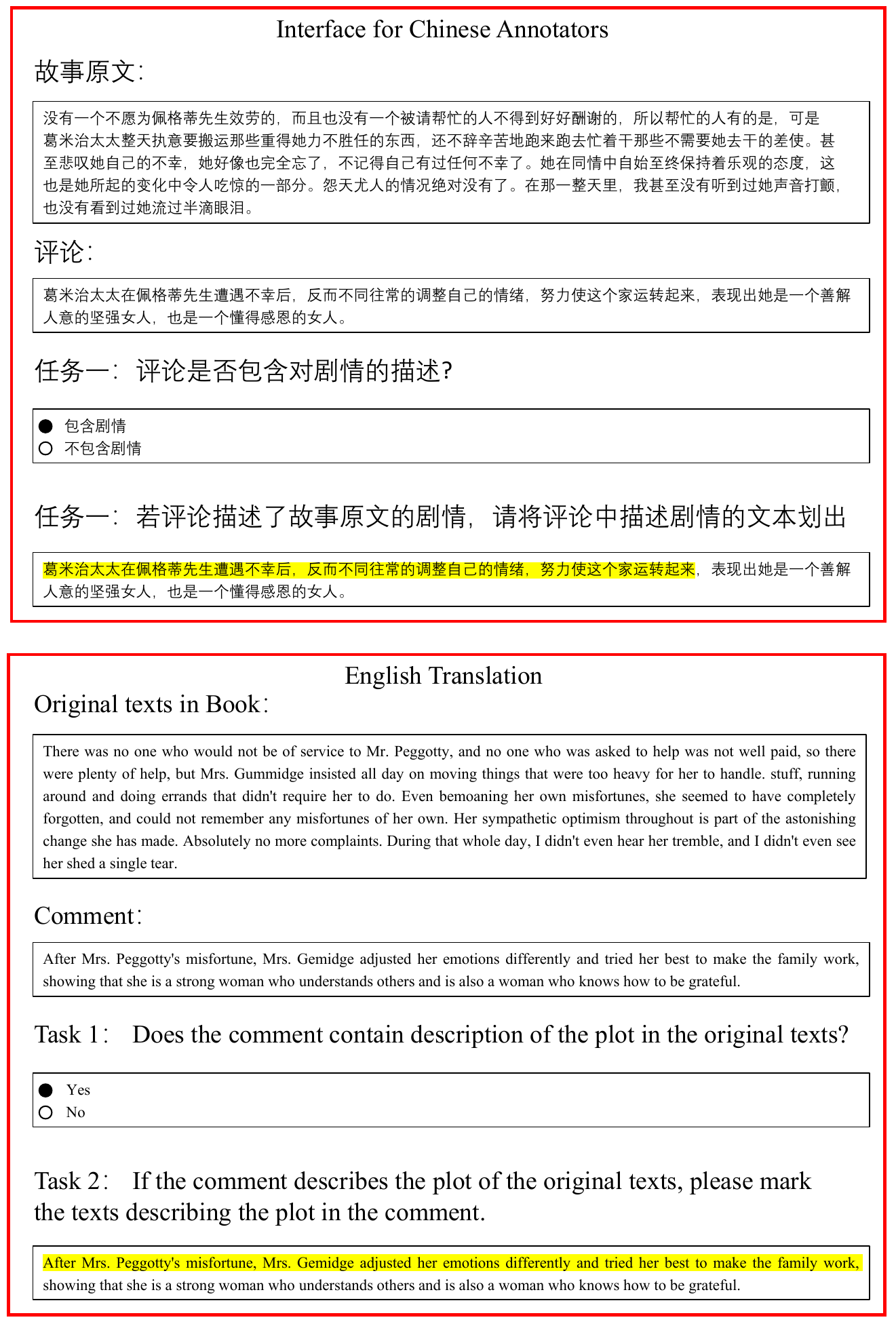} 
\caption{Interface for annotation.}
\label{interface}
\end{figure*}

Figure~\ref{interface} shows the interface for annotation. The original interface is in Chinese, we translate it into English for better reading.

\section{Details of \datasetname}

\subsection{Examples in \datasetname}
\begin{figure*}[h]
\centering
\includegraphics[width=\linewidth]{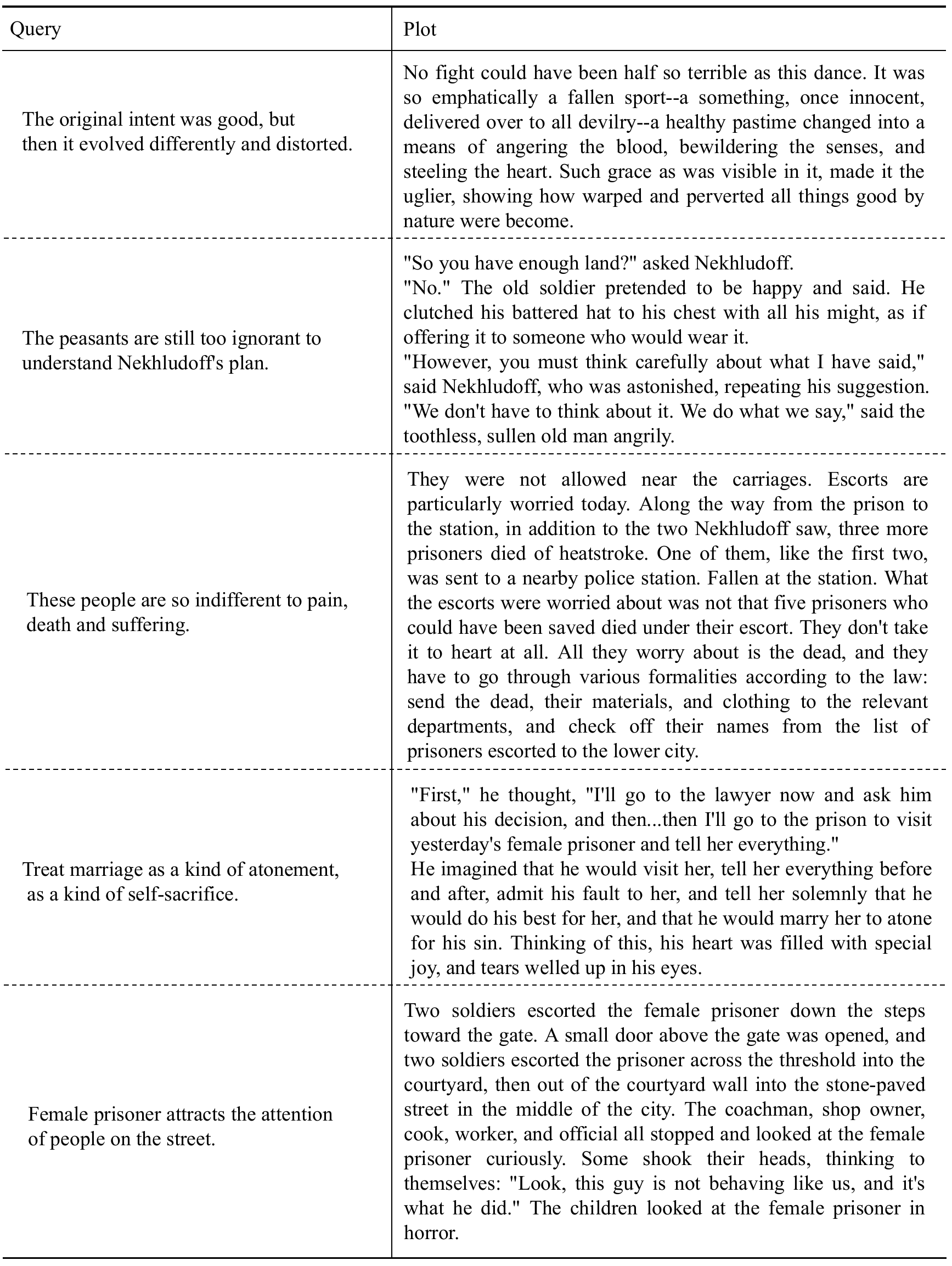} 
\caption{Examplse in \datasetname.}
\label{case1}
\end{figure*}

Figure~\ref{case1} shows some examples in \datasetnamens.

\subsection{Books in \datasetname} \label{corpus_detail}
Table~\ref{books} shows the book name in the corpus of test set and the number of queries and plot chunks for each book.

\begin{table*}[p]
\setlength\tabcolsep{10pt}
\centering
\begin{tabular}{lll}
        \toprule
        \bf Book Name &\bf \#Queries &\bf \#Plot Chunks \\ 
        \midrule
\emph{The Red and the Black} & 666 & 4353 \\
\emph{The Count of Monte Cristo} & 200 & 9013\\
\emph{The Adventures of Tom Sawyer Complete} & 121 & 1759\\
\emph{David Copperfield} & 153 & 6552\\
\emph{The Gadfly} & 134 & 2426\\
\emph{A Tale of Two Cities} & 325 & 2911\\
\emph{Crime and Punishment} & 404 & 5187\\
\emph{The Brothers Karamazov} & 217 & 8251\\
\emph{Les Miserables} &317 &12030 \\
\emph{Eugenie Grandet} &126 &1392\\
\emph{Tess of the d'Urbervilles} & 343 &3035\\
\emph{Notre-Dame de Paris} & 510 &4270 \\
\emph{The Call of the Wild} & 163 & 729\\
\emph{The Idiot} &122 &5480 \\
\emph{Moby Dick; or The Whale} & 125 &3429\\
\emph{Resurrection} &647 &3901\\
        \bottomrule
\end{tabular}
\caption{Books in the corpus of test set.}
\label{books}
\end{table*}

\section{Case Study}
\begin{figure*}[p]
\centering
\includegraphics[width=\linewidth]{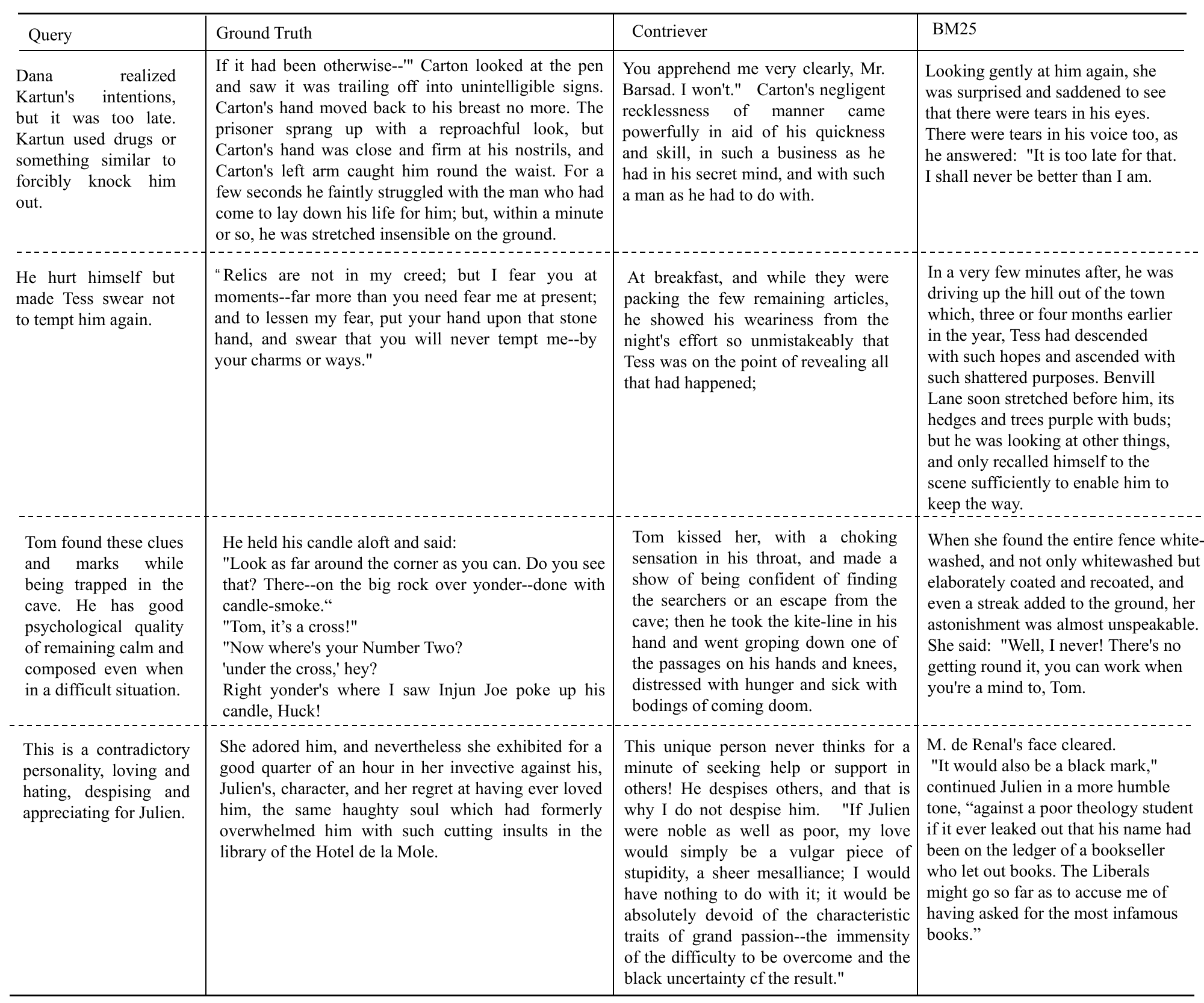} 
\caption{Comparison between ground truth and Top-1 results of Contriever and BM25.}
\label{cas_res}
\end{figure*}

Table~\ref{cas_res} shows the comparison of ground truth with Top-1 results retrieved by Contriever and BM25 respectively. The results of BM25 show that BM25 are limited to word overlap but cannot capture semantic level information. For the results of Contriever, they are limited to literal semantic matching, Contriever cannot deeply understand the meaning that the query really wants to express to find the most suitable plot.

\end{document}